\documentclass[twocolumn,reprint,superscriptaddress,showpacs,showkeys]{revtex4-1}
\usepackage[utf8]{inputenc}
\usepackage{lineno,hyperref}
\usepackage{cancel}
\usepackage{graphicx}
\usepackage{float}
\modulolinenumbers[5]
\usepackage{xcolor}
\usepackage{bm}
\usepackage{amsmath}
\usepackage{soul}
\newcommand\x{\times}
\newcommand\bigzero{\makebox(0,0){\text{\huge0}}}
\newcommand*{\bord}{\multicolumn{1}{c|}{}}

\begin{document}

\title{The Bellman equation and optimal local flipping strategies for kinetic Ising models}

\author{F. Caravelli}

\affiliation{  Theoretical Division (T4), Los Alamos National Laboratory, Los Alamos, New Mexico 87545, USA}

\begin{abstract}
There is a deep connection between thermodynamics, information and work extraction.  Ever since the birth of thermodynamics,  various types of Maxwell demons have been introduced in order to deepen our understanding of the second law. Thanks to them it has been shown that there is a deep connection between thermodynamics and information, and between information and work in a thermal system. In this paper we study the problem of energy extraction from a thermodynamic system satisfying detailed balance, from an agent with perfect information, e.g. that has an optimal strategy, given by the solution of the Bellman equation, in the context of Ising models. We call these agents kobolds, in contrast to Maxwell's demons which do not necessarily need to satisfy detailed balance. This in stark contrast with typical Monte Carlo algorithms, which choose an action at random at each time step. It is thus natural to compare the behavior of these kobolds to a Metropolis algorithm. For various Ising models, we study numerically and analytically the properties of the optimal strategies, showing that there is a transition in the behavior of the kobold as a function of the parameter characterizing its strategy.
\end{abstract}
\maketitle

\begin{figure}
    \includegraphics[scale=.6]{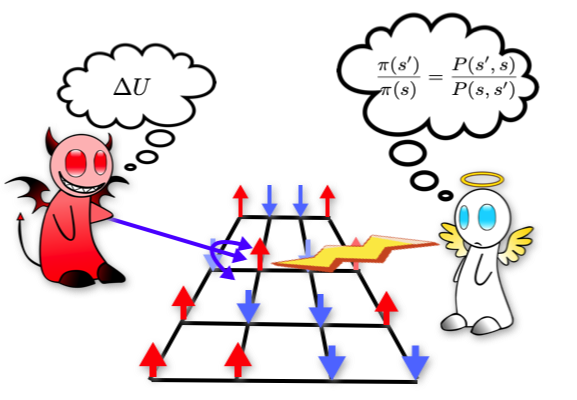}\\
    \caption{Representation of the game: the kobolds (a minor demon) observes the phase space of the Ising model and decides to attempt to flip a spin according to an optimal strategy in order to gain energy, attempting to locally lower the energy. However, detailed balance is enforced, and the spin flip can be rejected by the requirement of time reversibility.}
    \label{fig:pumuckle}
\end{figure}

\section{Introduction}
In later years there has been a lot of interest in the connection between thermodynamics, information and the role that thought experiments involving beings such as Maxwell demons play \cite{revmodphys,price,pekola,kosloff,brunner,plesch}. An example of a Maxwell demon \cite{maxwell} in a thermodynamic system is a being able to observe the speed of particles approaching a gate separating two gases, and open it if only the particle speed is such that the temperature of one of the gases can be raised, or acts in order for the pressure of one can be raised (temperature vs pressure demons). These thought experiments are useful to understand if the second law of thermodynamics can be locally violated. This line of thought has led later many researchers to actually propose various versions of the Maxwell demon, including notoriously Szilard \cite{szilard,lubkin}, Brillouin \cite{brillouin}, Landauer \cite{landauer} and Bennett \cite{bennett}. A typical Maxwell demon is able, quintessentially but with some restriction in certain cases, to do anything on the system. Depending on the point of view one takes, these thought experiments often can say something both about the system and the demon itself. If one assumes that thermodynamics is valid, then one can uses these arguments to infer what demons cannot do \cite{skordos}. On the other hand, if one assumes that the ability of the demons are valid, how thermodynamics can be violated. Either way, these thought experiments present a valid challenge for physicists, forcing them to think carefully about the nature of thermodynamic laws. As in the case of Landauer, and later Bekenstein and Hawking \cite{bekenstein1,bekenstein2,hawking}, these gedankexperiments with demons can lead to new discoveries about extreme regimes, ranging from nanoscale devices to black holes. A resolution of the violation of the second law can be obtained by assigning information to molecules or atoms, and introducing logical irreversibility on the operation of the operation performed by the demon, leading to an entropy increase due to information erasure. If however no information is erased and the operation is reversible, the second law is not violated. In fact, physical implementations of a Maxwell demon's have been proposed \cite{lloyd} and realized \cite{phot}, confirming such picture. These ideas can also be extended from the classical to the quantum realm, proving their extreme generality \cite{lande,peres,zurek}, and has also important applications to quantum thermodynamics \cite{delrio,horodecki} and non-equilibrium statistical mechanics \cite{jarzynski}.

Maxwell demons are often very far from real beings, in the sense that they stretch and bend slightly the roles of the games, and time is not as important. Innovation in technology is however a constant struggle between satisfying physical laws while gaining as much as possible from the system of interest, whether this is energy or computing power, or something else.
In this spirit, we consider here more mundane types of Maxwell demons, those which play by physical laws but with optimal strategies in complex landscapes. An optimal strategy is one which, given any  location of the agent in a state space of a system, performs an action which maximizes a certain function. These optimal strategies however require perfect ``information", e.g. the demon knows everything about the phase space of the system beforehand, and in particular has full phase space observability. We will use, for this purpose, the complex landscape of an Ising model. We note that this is not information as it is commonly stated in the theory of communication, but in the game theoretic sense. This simply means that the Kobold has perfect knowledge of energy landscape in the phase space, but has to satisfy detailed balance when performing a move; it thus follows a Markov decision process.

In particular, this paper is concerned  with the following problem. Consider a thermodynamic demon which knows everything about the dynamics of an Ising model, and desires to reach the ground state of the system by acting locally on a certain spin, while at fixed temperature. We choose the demon to act locally to closely represent the behavior of a physical device, or a Monte Carlo algorithm. For this reason, we introduce a kinetics for the Ising system which preserves detailed balance and thus microscopic reversibility. 
The question we then ask is: how hard would it be to act optimally, and to navigate through the phase space of the model in order to reach one of the states with minimum energy, while satisfying the detailed balance? Such entity is not quite as powerful as a Maxwell demon, as it still has to obey the laws of detailed balance and can only act locally on a single spin. Thus it resembles more of a kobold of german folklore, or goblin in english culture rather than a full fledged demon. However, the kobold has a perfect knowledge of the state space of the model (perfect information). A cartoon representation of the game is shown in Fig. \ref{fig:pumuckle}. In addition, the kobold values time, and has an associated parameter with which discounting occurs; discounting  is associated to a loss of energy as a function of time, and is a parameter that can be changed by the kobold. While microscopic reversibility is preserved at any intermediate step, there is a particular state at which the kobold decides to not act anymore, and saves the energy stored. In this sense, eventually microscopic reversibility is broken by the presence of a ``no action" by the kobold.

To answer this question, we need to specify some further details. First, we assume that we have a statistical physics system, described by spins variables $s_i=\pm 1$, and an energy written in the form
\begin{equation}
    H=\sum_{ij} J_{ij} s_i s_j.
\end{equation}

By navigating, we mean that we are able to observe our current phase space state $\phi=[s_1,\cdots,s_n]$, and that we need to act \textit{locally} via a map $a:\phi\rightarrow \phi$, where the action $a$ can act only on one spin or none in order to gain energy from the system.

This setup is not different from the typical Markov Chain Monte Carlo,  used to  thermalize or anneal a statistical model. For instance, the Metropolis-Hastings \cite{metropolis,hastings} or Glauber \cite{glauber} algorithm for the time evolution of a spin model: in the case of the latter the transition rates are determined by a spin flip. The model dynamics is  given by the spin flip probability
\begin{equation}
    W_{ij}=p(\phi_i\rightarrow \phi_j)=\frac{e^{-\beta \Delta U(\phi_i,\phi_j)}}{1+e^{-\beta \Delta U(\phi_i,\phi_j)}}
    \label{eq:glauber}
\end{equation}
implying that the system satisfies detailed balance, and can thus approach thermodynamic equilibrium \cite{onsager}. Then, we have an external clock with a discrete time which labels the operations and discounting.

Let us note that Glauber dynamics preserves the entropy of the model. This is because it the probabilty distributions and transition rate satisfy $P_i W_{ij} =P_j W_{ji}$, from which we obtain that the Schnakenberg entropy production \cite{schnakenberg} is zero, e.g.
\begin{equation}
    \Pi(t)=\frac{k}{2}\sum_{ij}(P_i(t) W_{ij} -P_j(t) W_{ji})\log\frac{P_i(t) W_{ij}}{P_j(t) W_{ji}}=0.
\end{equation}
Thus, the change in free energy of the model reduces to the change in potential energy of the system. Interestingly, this also means that a kobold acts as a reversible computer \cite{frank}. Another point of view is that the system represents the battery, and the kobold is an operation by which we wish the extract energy from the system without wasting it in entropy production.

One question one might ask is why is a kobold different from a standard Monte Carlo ``demon". We could in fact consider a system out of equilibrium and let it thermalize, end extract the energy as a result of thermalization. However, if $\bar E$ is the expected thermalization energy, if $E(0)<\bar E$ the demon would actually have to supply energy to the system. A kobold, on the other hand, has an optimal strategy for any initial condition, and can act on its discounting factor to try to maximize it. In doing so, it can also reduce fluctuations over the gained energy.

The paper is organized as follows. We will first discuss the theoretical underpinning of kobolds, envisaged as a stochastic Bellman equation for the optimal strategy \cite{bellman1,bellman2,reinfl}. We will derive an analytical formulation for the Bellman equation restricted to the case of Glauber dynamics, and show that this can be written in terms of projector operators on the phase space. We will then provide numerical results both for the optimal strategies and discounted gained energy.
In particular, we will show that depending on the discounting factor, kobold can take a greedy or a wise strategy approach; a greedy strategy is one in which the local spin flip action is such that it always minimizes the energy, while a wise strategy allows for local increases in energy. In this respect, a greedy approach can be compared to a low temperature Metropolis algorithm, while a wise approach to a higher temperature one. However, the analogy ends there, as the kobold tries independently to maximize the energy extracted from the system. This is the reason why we compare the behavior of such an agent to a ``Metropolis" agent.

Conclusions follow.

\section{Optimal strategies and the Bellman equation}
\subsection{The discounted reward}
We assume the following rules for the abilities and limitations of the kobold. First, the kobold knows everything about the system: the couplings, a complete picture of the phase space, and thus also the location of the ground state. In the language of economics and game theory, the kobold has perfect information (rather than imperfect, which would occur if the kobold only knows a certain part of the phase space). However, it has some restrictions, most importantly it has to respect constraints such as the detailed balance. This implies that while it might decide to attempt a spin flip to gain energy, whether this occurs or not is determined by the acceptance rule at the temperature of system. The second restriction is that the kobold can only operate locally in time and space, meaning that it has to perform actions sequentially, and one spin at the time, but any spin of choice. The third restriction is that the kobold does not get to choose the initial state from which it starts to operate on the system. What the kobold gains by playing this game is energy. If at each time step the kobold gains $\Delta U_t$ of energy via a certain action (spin flips), after $T$ steps the kobold will have gained
\begin{equation}
    R_{\gamma}=\sum_{t=1}^\tau R_\gamma(t)=\sum_{t=1}^\tau \gamma^t \Delta U(\phi_t,\phi_{t-1});
    \label{eq:discr}
\end{equation}
in the equation above $\gamma\in (0,1]$ is a discounting factor: the kobold knows the rules of finance, and knows that energy today is better than energy tomorrow. 

In practice, if $\gamma<\gamma^\prime<1$, this means that even if the actions are such that the energy is positive at each time step one has $R_{\gamma}<R_{\gamma^\prime}$. When $\gamma=1$, we simply have
\begin{equation}
    R_1=\sum_{t=1}^\tau R_1(t)=\sum_{t=1}^\tau (U_t-U_{t-1})=U_{\tau}-U_0.
\end{equation}
In the equation above, and in the context of the Bellman equation, the parameter $\gamma$ characterizes the kobold strategy, as it defines how important is time to the kobold.

\subsection{Bellman equation}

In general, a kobold agent wants to maximize the reward over time (possibly an infinite horizon) via local actions $a_\phi=\pi(\phi)$, meaning that by observing the local state of the system, the kobold can attempt to flip one spin at time $t$ with the intention of gaining energy. Thus, the parameter $
\gamma$ represents how important is the speed at which the kobold attempts to reach the ground state. It is important to stress that the form of the discounted energy of eqn. (\ref{eq:discr}) can be obtained by assumptions on the time invariance of the optimal strategy, and is thus natural.

In the following, we use a notation in which $\pi(\phi)$ determines both an action and a state, as the two in this context are the same. If the action $\pi(\phi)$ is to flip the k-th spin of $\phi$, then this determines a new state $\phi^\prime$. Of course, if the kobold reaches the ground state, it will want to perform no action. This implies that if the system contains $N$ spins, there can be $N+1$ possible actions. We say attempt because of course the acceptance probability of such action is the determined by the system's temperature, and thus if the currently the kobold is in state $\phi$, the action $\pi(\phi)$ of the kobold can lead to a state $\phi^\prime$ with probability $P_T(\phi; \pi(\phi),\phi^\prime)$, with $T$ being the temperature.
The average energy reward is then given by
\begin{eqnarray}
    \langle R_\gamma(t)\rangle \equiv \langle U\rangle_{t,\pi(\phi)}=\sum_{\phi^\prime} P_T(\phi; \pi(\phi),\phi^\prime) \gamma^t \Delta U(\phi_t,\phi_{t-1}).\nonumber
\end{eqnarray}

For the kobold, an ideal world would be such that $P_T(\phi; \pi(\phi),\phi^\prime)=\delta_{\phi,\pi(\phi)}$. The setup is now such that the kobold is playing a (reversible) Markov Decision Process in discrete time. A Monte Carlo algorithm plays typically the same game, with the difference that the action $\pi(\phi)$ is random in some form, e.g. a random single (Metropolis or Glauber) or multiple (Kawasaki) spin flip \cite{eli}. The kobold instead plays like an economist: uses dynamic programming and the notion of discounting, knows everything about the system and applies an optimal strategy $\pi^*(\phi)$.
Under these assumptions, an infinite-horizon decision problem takes the form of a maximization, e.g.
\begin{eqnarray}
    V^*_{T,\gamma}(\phi)=\text{max}_{a_0,\cdots,a_\tau} \langle R_\gamma(a_0,\cdots, a_\tau)\rangle,
\end{eqnarray}
where $\phi$ is the starting state and $a_i$ the actions taken at each time step.  The optimal reward $V_{T,\gamma}^*(\phi)$ is the maximum (discounted) reward that the kobold can obtain starting from a certain state in phase space. For $\gamma=1$, this  corresponds to the maximum energy the kobold can extract.
We then see the reason of such  gedankexperiment. The kobold represents the best possible algorithm designed to reach the ground state of the system, and then also represents the best possible line of action an algorithm can take at fixed temperature. Here, the assumption is that the system is \textit{not} annealed, probabilities are time independent and thus the kobold can act with an infinite time horizon. For this type of problems, the optimal solution $\pi^*(\phi)$, and the optimal reward $V_{T,\gamma}^*(\phi)$ can be obtained by solving the stochastic Bellman equation. Above, $\pi^*(\phi)$ is the best possible action that the kobold takes if it finds itself in state $\phi$. 
The Bellman equation is given by the linear relationship
\begin{equation}
    V_{T,\gamma}^*(\phi)=\sum_{\phi^\prime}P_T(\phi; \pi^*(\phi),\phi^\prime)\Big(-\Delta U(\phi,\phi^\prime)+\gamma V^*_{T,\gamma}(\phi^\prime) \Big).
    \label{eq:bellman}
\end{equation}

The equation above is written implicitly: to solve it one would have already to have $\pi^*$, which we do not. There are many ways to solve it however, and we use here the strategy iteration method which starts from a random initial state $V_0(\phi)$, and then uses the iteration 
\begin{eqnarray*}
    V_{T,\gamma; k+1}(\phi)&=&max_a\sum_{\phi^\prime}P_T(\phi; a,\phi^\prime)\Big(-\Delta U(\phi,\phi^\prime)\nonumber \\
    &&\hspace{3.5cm}
    +\gamma V_{T,\gamma;k}(\phi^\prime) \Big),
    \label{eq:bellmanstrategyit}\\
    \pi^*(\phi)&=&arg\ max_a\sum_{\phi^\prime}P_T(\phi; a,\phi^\prime)\Big(-\Delta U(\phi,\phi^\prime)\nonumber \\
    &&\hspace{3.8cm}+\gamma V^*_T(\phi^\prime) \Big).
    \label{eq:bellmanoptstrategyit}
\end{eqnarray*}
which gives, at convergence, the optimal strategy $\pi^*(\phi)$.

Note that here we immediately face a computational problem. While the number of actions is $N+1$, the number of states $\phi$ scales exponentially with the size of the system, and thus unlike the kobold we will have to use small systems by Monte Carlo standards. Despite such curse of dimensionality, the Bellman solution for optimal sequential Markov decision processes is regarded as the most feasible one \cite{reinfl}. Nonetheless, this will be sufficient to obtain a picture of the complexity of Ising models from the point of view of the kobold, as we will discuss in a moment.

\subsection{Application in kinetic statistical mechanics.}
Let us provide an immediate comment on why such technique is useful. Since we are essentially finding the functions $V^*$ and $\pi^*$ for every state of the model, if one is interested only in the energy of the model, exhaustive search (or brute force) works much better than solving the Bellman equation. However,  Bellman's optimal strategy and discounted energy give a more complete picture of how the state space of an Ising model is tangled, and how one could unentangle it and navigate through the states, or make less blind moves in Monte Carlo algorithms.

First, we note that eqn. (\ref{eq:bellman}) can be written in an explicit form using the Glauber transition of eqn. (\ref{eq:glauber}), given a certain action $\pi$ on the phase space:
\begin{eqnarray}
    P_T(\phi; \pi(\phi),\phi^\prime)=\frac{\delta_{\phi \phi^\prime}+\delta_{\phi^\prime \pi(\phi)}e^{-\beta \Delta U(\phi,\pi(\phi))}}{1+e^{-\beta \Delta U(\phi,\pi(\phi))}}
    \end{eqnarray}
    
from which it follows that we can write the Bellman equation in the form
\begin{equation}
    \sum_{\phi^\prime}\mathcal O(\phi,\phi^\prime)V_{T,\gamma}^*(\phi^\prime)=-\frac{\Delta U\big(\phi,\pi^*(\phi)\big)}{1+e^{-\beta \Delta U(\phi,\pi^*(\phi))}}
    \label{eq:bellman2}
\end{equation}
where 
\begin{eqnarray}
    \mathcal O(\phi,\phi^\prime)&=&\delta_{\phi \phi^\prime}\frac{1+e^{-\beta \Delta U(\phi,\pi^*(\phi))}-\gamma}{1+e^{-\beta \Delta U(\phi,\pi^*(\phi))}}\\
    &-&P_{\pi^*}\big(\phi,\phi^\prime\big)\frac{\gamma e^{-\beta \Delta U(\phi,\pi^*(\phi))}}{1+e^{-\beta \Delta U(\phi,\pi^*(\phi))}}
\end{eqnarray}
Above, $P_{\pi^*}\big(\phi,\phi^\prime\big)$ is a state transition matrix, e.g. $P_{\pi^*}\big(\phi,\phi^\prime\big)=\delta_{\phi^\prime,\pi^*(\phi)}$. After a brief calculation, it follows that we can write the exact solution for $V^*$ in the form:
\begin{eqnarray}
    V^*_{T,\gamma}(\phi)=\sum_{\phi^\prime}(I-\gamma D P_{\pi^*})^{-1}_{\phi,\phi^\prime}   \tilde U(\phi^\prime) .
\end{eqnarray}
where 
\begin{eqnarray}
    \tilde U(\phi)&=&-\Delta U(\phi,\pi^*(\phi))R(\phi),\ \ \
    D(\phi,\phi^\prime)=\delta_{\phi,\phi^\prime}R(\phi)\nonumber \\
    &&\text{and where  }R(\phi)=\frac{e^{-\beta \Delta U(\phi,\pi^*(\phi))}}{1-\gamma+e^{-\beta \Delta U(\phi,\pi^*(\phi))}}.\ \ \ \ \ \ 
\end{eqnarray}
 Assuming a random initial condition, the average maximum discounted energy obtained from the kobold is given by 
\begin{equation}
    \langle V_{T,\gamma}\rangle=\frac{1}{2^N}\sum_{\phi} V^*_{T,\gamma}(\phi),
\end{equation}
where $N$ is the number of spins. This is the average (discounted) energy that an optimal player (a kobold) can achieve by acting locally on the system, and against the temperature, by reaching the ground state. In general, the Bellman equation selects one possible action for each phase space state, but there might be some other stochastic policies, which we do not consider here, such that $\pi$ is a stochastic function as well. This helps however in evaluating the amount of information that a policy contains. If we have $M$ possible actions per phase state, and $K$ states in phase space, then we have $M^K$ possible policies. We can then evaluate the information in the policy using the entropy (in a base 2) $S_\pi=K \log_2 M$, which is an estimation of the amount of information associated to a kobold, including an optimal one.

\subsection{Example: classical Zener Hamiltonian.} Let us consider first a simple application of the equations above. We take as a model the classical Zener Hamiltonian for an Ising spin in an external field, which is given by $H=\mu_0 h s$, with $s=\pm 1$ representing a single spin. In this case we have only two states, and the optimal strategy is easy to guess: it is a spin flip if $sign(s)=-sign(h)$, and do nothing otherwise. Let us assume $h>0$. Then,
\begin{eqnarray}
\begin{pmatrix}
V_{T,\gamma}^*(+)\\
V_{T,\gamma}^*(-)
\end{pmatrix}&=&\begin{pmatrix}
\frac{2-\gamma}{2} -\frac{\gamma}{1+\gamma}& 0\\
\frac{\gamma e^{\beta \mu_0 h}}{1+e^{\beta \mu_0 h}} & \frac{1+e^{\beta \mu_0 h}-\gamma}{1+e^{\beta \mu_0 h}}
\end{pmatrix}^{-1}
\begin{pmatrix}
0\\
\mu_0 h
\end{pmatrix}\nonumber \\
&=&\frac{1}{d}\begin{pmatrix}
\frac{1+e^{\beta \mu_0 h}-\gamma}{1+e^{\beta \mu_0 h}} & - \frac{\gamma e^{\beta \mu_0 h}}{1+e^{\beta \mu_0 h}}\\
0 &\frac{2-\gamma-\gamma^2}{2+2\gamma}
\end{pmatrix}
\begin{pmatrix}
0\\
\mu_0 h
\end{pmatrix}\nonumber \\
&=&\frac{\mu_0 h}{1+e^{\beta \mu_0 h}-\gamma}\begin{pmatrix}
-(\gamma e^{\beta \mu_0 h})\frac{2+2\gamma}{2-\gamma-\gamma^2}\\
1+e^{\beta \mu_0 h}
\end{pmatrix}\nonumber
\end{eqnarray}
where $d=\frac{2-\gamma-\gamma^2}{2+2\gamma}\frac{1+e^{\beta \mu_0 h}-\gamma}{1+e^{\beta \mu_0 h}}$.
It follows that the average energy the kobold can gain is
\begin{eqnarray}
    \langle V^*_{T,\gamma}\rangle=\frac{\mu_0 h}{2(1+e^{\beta \mu_0 h}-\gamma)}\big(1+e^{\beta \mu_0h}(1-2\frac{\gamma+\gamma^2}{2-\gamma-\gamma^2})\big)\nonumber
\end{eqnarray}
It follows that with a proper discounting strategy, e.g. $\gamma\in[0,\frac{1}{2}(\sqrt{3}-1)]$, the Kobold can always extract energy from the system, at any temperature.

\subsection{Properties of $P_{\pi^*}$} 
In this section we discuss the properties of $P_{\pi^*}$, which are useful to understand the behavior of a kobold in the case of an Ising system via local spin flipping.

In fact, as it turns out, the matrix $P_{\pi^*}(\phi,\phi^\prime)$ is a projector operator, as its spectrum is only composed of $0$'s and $1$'s. Intuitively, the proof follows from the fact that we have a directed graph of outdegree equal to one, with an absorbing state, which is what we discuss below.

\begin{figure}
    \centering
    \includegraphics[scale=0.36]{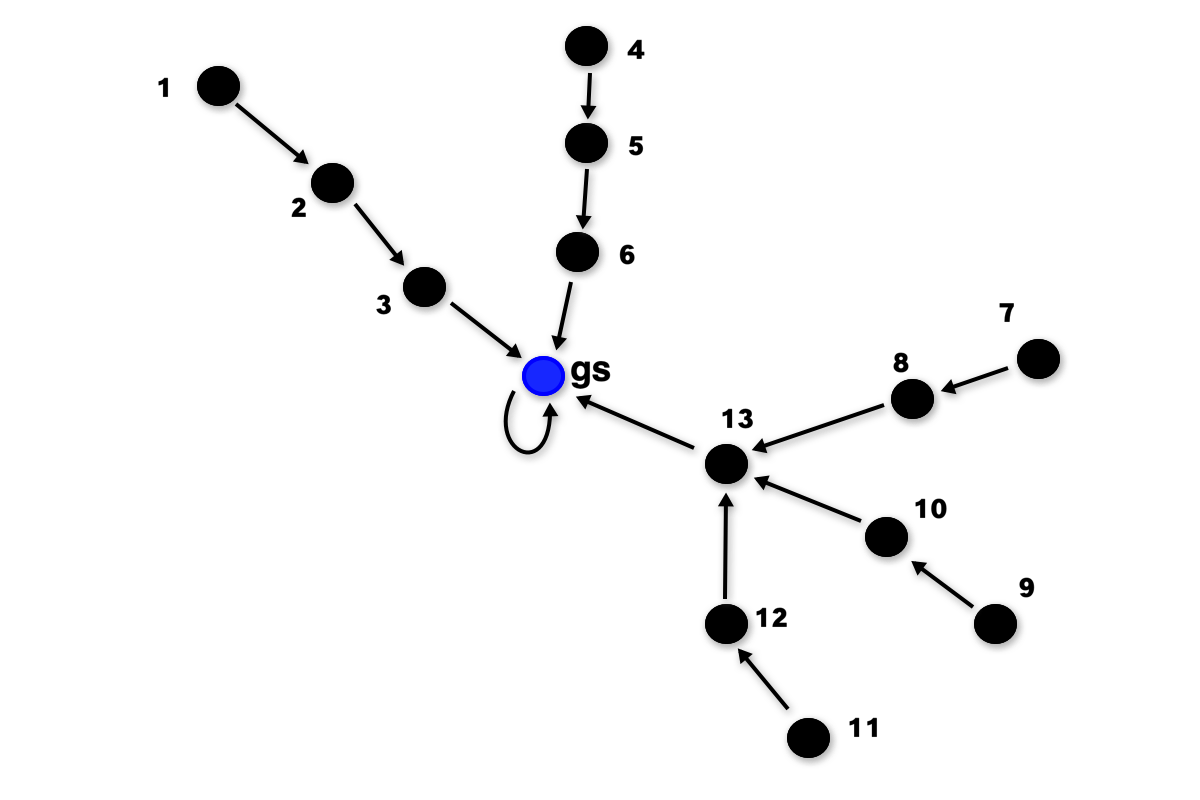}
    \caption{The graph representation of the an optimal strategy $\pi^*$ restricted to the basin of attraction of the absorbing state. Since the graph is directed and acyclic, there is a node labeling $l(i)$ such that $l(i)>l(i^\prime)$ if $i^\prime$ can be reached by a directed path from $i$.}
    \label{fig:proj}
\end{figure}

The important quantity in the Bellman equation for kinetic Ising models is the transition matrix $P_{\pi^*}$, which could be in principle a permutation matrix. However, as it turns out, the optimal strategy is always such that $P_{\pi^*}$ is a projector operator, e.g. $P_{\pi^*}^2=P_{\pi^*}$, as we will prove in a moment.  The kobold's strategy is to lower the energy, by transitioning between states and eventually to one of the absorbing states. 
Then, the basins of attraction of a particular absorbing state $G$ as a sink in a directed tree, similar to Fig. \ref{fig:proj}, which follows from the fact that the outdegree of every node is always one. 
If we have $G$ absorbing states, then the matrix $P_{\pi^*}$ can be written in block diagonal form. Let us call these sub-blocks $P^G_{\pi*}$; then, simply one has $P_{\pi^*}=\bigoplus_{G} P^G_{\pi*}$ and then we can focus on each sub-block. Now, we can label the nodes such that if $i$ is the numerical value of a certain node and $i^\prime$ down the tree, then $l(i)>l(i^\prime)$. Such labeling is always possible because we have a directed acyclic graph, and one example is shown in Fig. \ref{fig:proj}. The fact that it is acyclic follows from the fact that we have an absorbing state for a graph with outdegree one. In fact, assume by absurd that one has a cycle in such a graph with outdegree one. If it is a cycle, there cannot be an absorbing state, and thus there must be at least one node with degree three and with outdegree two, since one directed edge goes inside the cycle and the other in the direction of the absorbing state. However, this is incompatible with the fact that the outdegree must always be one.
With the labeling $l$ we see that these energy state transitions correspond to an upper triangular matrix elements of $P^G_{\pi^*}$. Since the absorbing state is the only state which has null action (the kobold will want to remain in that state), this means the absorbing state state is the only element with a one on the diagonal. 
Each sub-block can always be written, via the state relabeling, as
\begin{eqnarray}
    P^G_{\pi^*}=
    \begin{pmatrix}
    0    & \x       & \x    & \x    & \x \\ \cline{1-1}
    \bord & 0       & \x    & \x    & \x \\ \cline{2-2}
          & \bord    & 0    & \x    & \x \\ \cline{3-3}
          & \bigzero & \bord & 0    & \x \\ \cline{4-4}
          &          &       & \bord & 1 \\ \cline{5-5}
  \end{pmatrix}
\end{eqnarray}
from which it follows, since the matrix is triangular, that if such sub-block is $D$ dimensional, then $D-1$ eigenvalues are $0$, and only one is $1$. This is enough to show it is is a projector operator, and that it satisfies the condition $(P_{\pi^*}^G)^2=P_{\pi^*}^G$. Then, the spectrum of $P_{\pi^*}$ is simply determined by the number of absorbing states. In fact, the number of $1$s correspond to the number of absorbing states of the system, and the rest of the spectrum contains only zeros. Since every diagonal sub-block is a projector operator, so is $P_{\pi^*}$.

This result is useful for the following reason. First, in the limit $\gamma\rightarrow 1$, we have $D\rightarrow I$ and $R(\phi)\rightarrow 1$. We can write explicitly the inverse, in the neighborhood of $\gamma=1^-$ and using the property that $P^2_{\pi^*}=P_{\pi^*}$, as
\begin{eqnarray}
    (I-\gamma P_{\pi^*})^{-1}=I+\frac{\gamma}{1-\gamma} P_{\pi^*},
\end{eqnarray}
and then the discounted energy values are simply given by
\begin{eqnarray}
    V^*_{T,\gamma\approx 1}(\phi)
    &=& -\sum_{\phi^\prime}\Big(I+\frac{\gamma}{1-\gamma}  P_{\pi^*}\Big)_{\phi,\phi^\prime}\Delta U(\phi^\prime,\pi^*(\phi^\prime)) \nonumber \\
\end{eqnarray}
We can thus use the exact inverse to regularize the limit $\gamma\rightarrow 1$, which is otherwise ill-defined in the general case. We can use the following assumption. Typically, $\partial_\gamma \pi^{*}=0$ almost everywhere. Then, assuming that the optimal strategy can analytically extended from $\gamma=1-\epsilon$ to $\gamma=1^{-}$, i.e. that it is constant, we can obtain a discounted effective energy of the form
\begin{eqnarray}
    \tilde V^*_{T,\gamma}(\phi)&=&\lim_{\gamma\rightarrow 1^{-}}(1-\gamma)  V^*_{T,\gamma}(\phi) \nonumber \\
    &=&\sum_{\phi^\prime}  (P_{\pi^*})_{\phi,\phi^\prime}\Delta U(\phi^\prime,\pi^*(\phi^\prime)),\nonumber
\end{eqnarray}
which is a regularized $\gamma=1$ limit for the optimal strategy. In this regime, all absorbing states are ground states. 

In fact, we have been careful in calling $G$ the absorbing state and not the ground state. If $\gamma=1$, of course all absorbing states are ground states, but at $\gamma<1$ this is not guaranteed, although likely for $\gamma\approx 1$. Intuitively, this is because the kobold might find more rewarding to stop at a certain state rather than attempting to reach the ground state, given that there is a cost in how long the game takes, and in the intermediate steps that might actually increase rather than lowering the energy.

\section{Numerical results}
We consider, for the purpose of this paper, four types of Ising models, of which three ferromagnetic and one frustrated.  The testbed of our analysis are the Ising models given below 
\begin{equation}
    H=\begin{cases}
    \frac{1}{2}\sum_{i=1}^{N-1} s_i s_{i+1} &\ \ \ \ \  \text{Ising 1D}, \\
    \frac{1}{2}\sum_{<ij>=1}^{N_1 N_2} s_i s_{j} &\ \ \ \ \  \text{Ising 2D}, \\
\frac{1}{2N}\sum_{ij=1}^Ns_i s_j &\ \ \ \ \  \text{Curie-Wei}\ss, \\
\frac{1}{2\sqrt{N}}\sum_{ij=1}^NJ_{ij}s_i s_j &\ \ \ \ \ \text{Spin Glass} \ \ \ J_{ij}=\pm 1.
    \end{cases}
\end{equation}
where in the latter case we consider $P(J_{ij}=\pm)=1/2$.
Since solving the Bellman equation requires solving iteratively a vectorial equation of the size of the phase space, we are forced to study relatively small systems, with $12$ spins.

\begin{figure}
    \centering
    \includegraphics[scale=0.13]{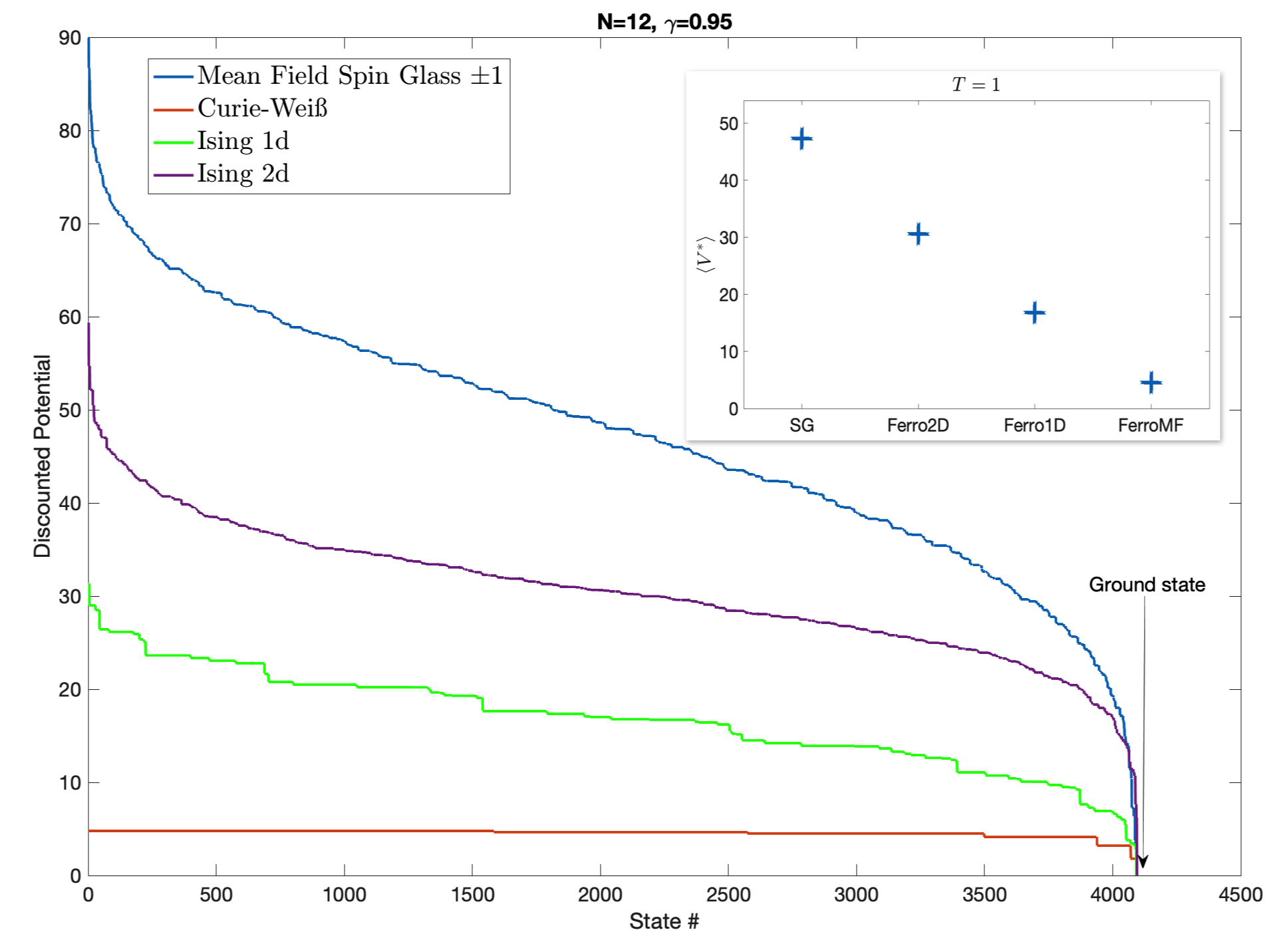} \includegraphics[scale=0.2]{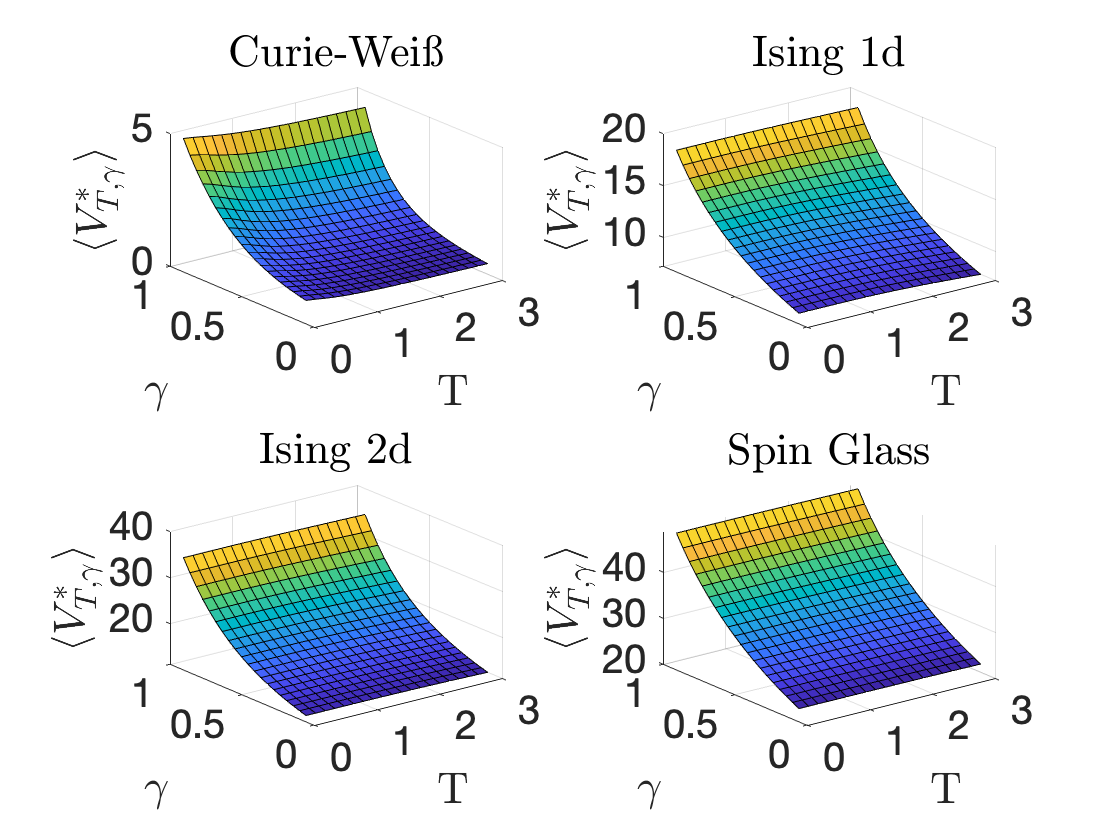}
    \caption{\textit{Top:} maximum discounted energy $V^*_{T,\gamma}$ as a function of the states $\phi$ for $N=12$, ordered from high to low, and $\langle V^*_{T,\gamma}\rangle$ shown in the inset. We see the difference between the spin glass, Ising 2D and Ising 1D and the Curie-Wei\ss\ model. Since $\gamma=0.95$, the low energy states are associated to ground states of the model. \textit{Bottom:} The profile of the function $\langle V_{T,\gamma}^*\rangle$ as a function of $T$ and $\gamma$. While the dependence on $T$ is negligible for these values of $T$, the dependence on $\gamma$ is strong, due to the fact that high discounting rates imply less value to the energy obtained in the future.}
    \label{fig:discounted}
\end{figure}

\begin{figure}[ht!]
    \centering
    \includegraphics[scale=0.18]{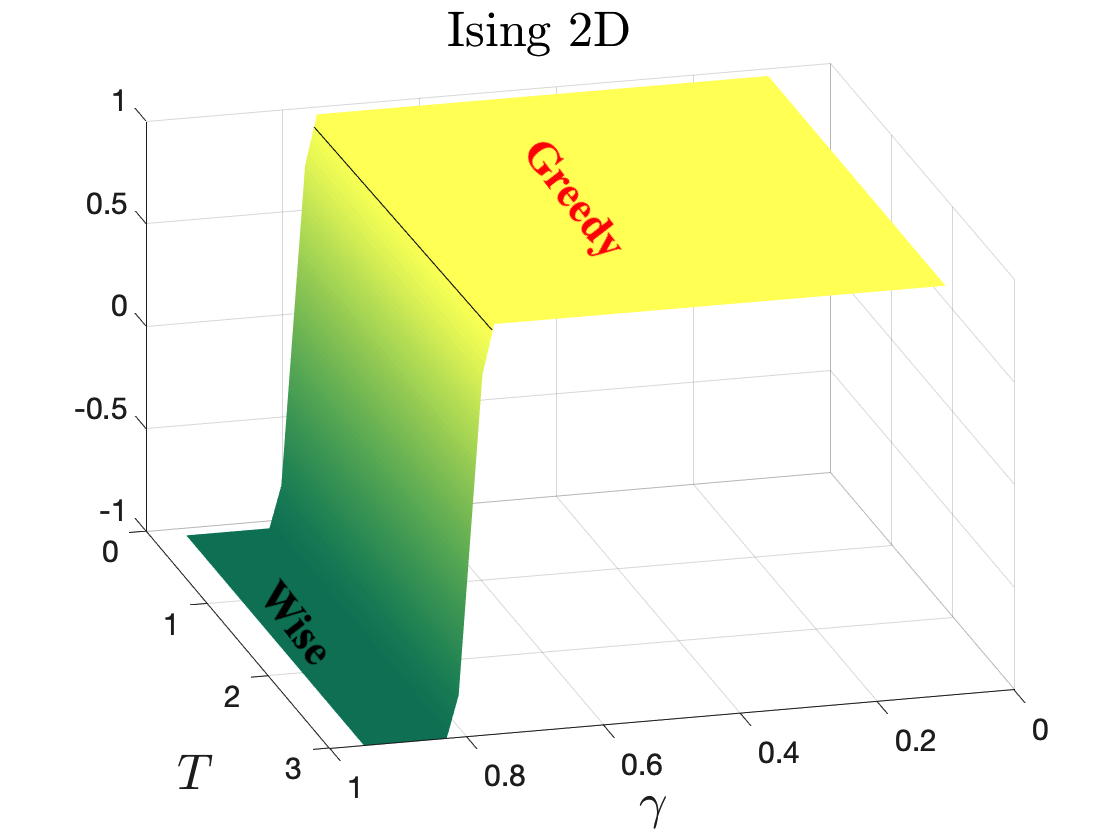}\\
        \includegraphics[scale=0.18]{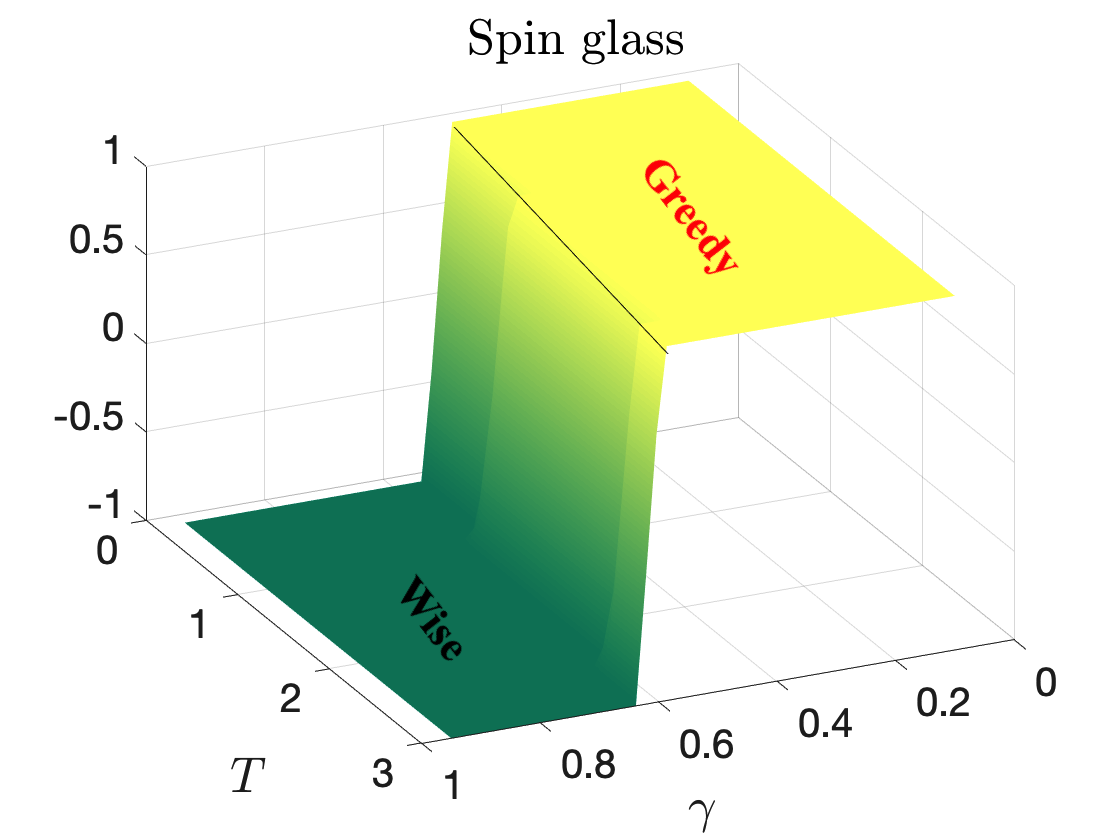}
        \caption{ Change in kobold strategy as a function of temperature and discounting rate. A greedy strategy implies only lowering the energy at every step, while a wise one can also accept intermediate energy increases. }
            \label{fig:greedyvswise}
\end{figure}
\begin{figure*}[ht!]\centering
        \includegraphics[scale=0.18]{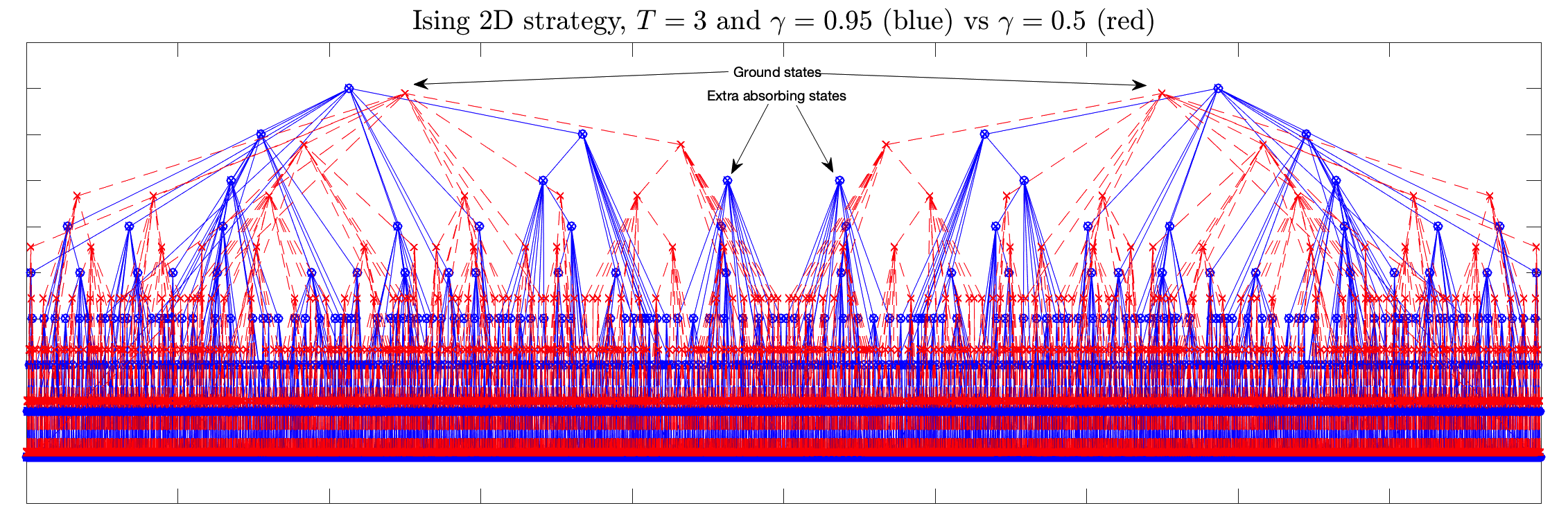} 
    \includegraphics[scale=0.18]{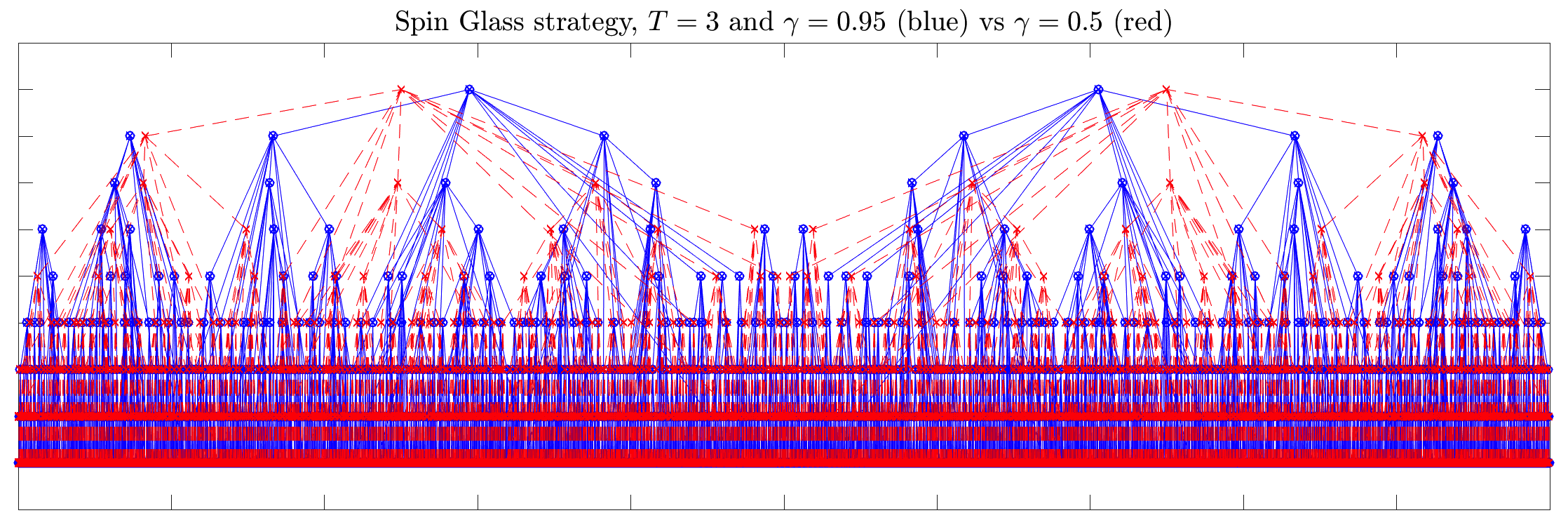}
    \caption{Kobold strategy tree (top are absorbing states) as a function of the $\gamma$ for $T=3$ for the Ising 2D and spin glass cases. Red curves are low discounting, while blue are high discounting rates. We see the emergence of two extra absorbing states on top of the Ising 2D ground states.}
    \label{fig:tree}
\end{figure*}

We iterate eqn. (\ref{eq:bellmanstrategyit}) until $\frac{1}{2^N}\|V_{k+1}-V_{k+1}\|^2<\epsilon$, with $\epsilon=10^{-5}$, read the strategy out and sort the spin states with descending discounted potential $V^*(\phi)_{T,\gamma}$. The plots are shown in Fig. \ref{fig:discounted} for $\gamma=0.95$. In the inset of Fig. \ref{fig:discounted}, we plot $\langle V_{T,\gamma}^*\rangle$
for each model at $\gamma=0.95$. The quantity $\langle V_{T,\gamma}^*\rangle$  is the average utility (discounted energy) that the kobold can obtain from the Ising model, assuming that we let him start from any spin state at at infinite time. Such average shows, in arbitrary units of energy and assuming all being equal, which model lead to more gains via a Monte Carlo method,  withing the assumption of only local moves. Intuitively, we see that the spin glass lead to more gains than the others, with the Ising 2D being harder than Ising 1D model, and the mean field ferromagnetic model being the easiest in comparison. Such hierarchy seemingly makes sense from a computational perspective, as we would expect the spin glass to have a longer way down to the ground state. However, we note that the two key parameters we need analyze are $\langle V_{T,\gamma}^*\rangle$ as a function of temperature, of course, and as a function of the parameter $\gamma$. If $\gamma\approx 1$, time for the kobold is not an issue.
At strong discounting, however, longer chains to reach the ground state imply losses. This can be seen in Fig. \ref{fig:discounted} (bottom), in which we see that for all models at stronger discounting ($\gamma\rightarrow 0$) the effective value is reduced.

An interesting question is how the optimal strategy of the kobold changes as a function of $T$ and $\gamma$. We explore this in two ways. First, we analyze whether the only allowed strategies are those that go down in energy, which we call ``greedy" kobold strategies ($S=+1$). On the other hand, if the kobold uses strategies such that at a certain point it actually increases momentarily the energy, these are "wise" ($S=-1$), as they allow the kobold to reach a lower state faster. We find that for the 1D Ising model and the Curie-Wei\ss\ model the strategy is always greedy. However, we plot $S$ as a function of $\gamma$ and $T$ in Fig. \ref{fig:greedyvswise}. We see that for the 2D Ising model, the parameter $S$ does not depend on $T$, but it depends on $\gamma$ and for $\gamma>0.8$, the strategy becomes \textit{wise}, while it is \textit{greedy} for $\gamma<0.8$. For the spin glass case, instead, we find a dependence on the temperature as well, with a transition between a greedy and wise strategy at approximately $\gamma=0.62$ for $T=3$, and $\gamma\approx 0.55$ for $T=0.1$.

 The maximum number of steps required (on average) can be estimated by the depth of the optimal policy tree. For the cases of the Ising 2D and Spin Glasses, these are shown in Fig. \ref{fig:tree}, showing the difference between a high and low discounting rates for $T=3$. As we see from the change of the trees, shown in different colors, the kobolds uses a different strategy at different discounting rates. Interestingly, at high discounting rates the kobolds develops two extra absorbing states, on top of the ground states of the model for the Ising 2D case.
 
\begin{figure}
    \centering
    \includegraphics[scale=0.18]{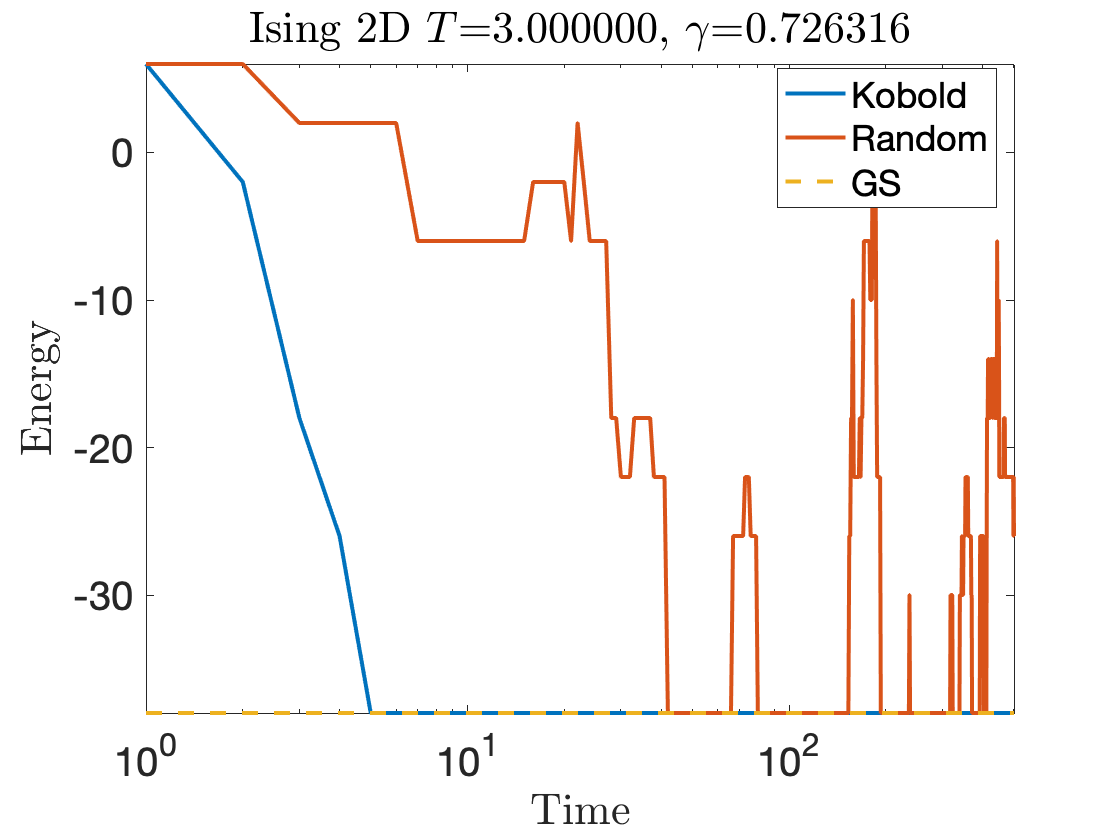}\\
    \includegraphics[scale=0.18]{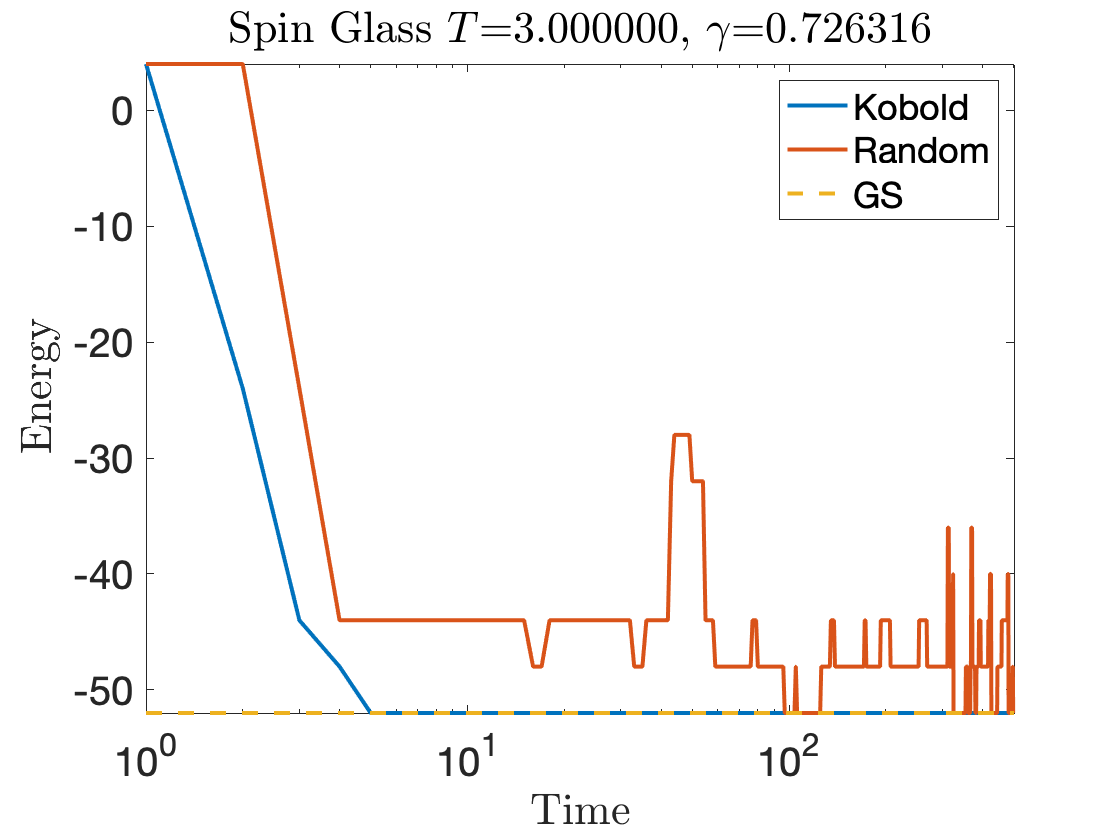}
    \caption{Example of Markov Chain Monte Carlo for the Ising and Spin Glass examples, for $N=12$. We compare the dynamics to the Ising ground state of the associated model.}
    \label{fig:MCvsK}
\end{figure}

 We can see the difference between the kobold strategy and the Monte Carlo thermalization process in Fig. \ref{fig:MCvsK} for the case of the Ising 2D and Spin Glass. The random strategy fluctuates strongly in energy, while the kobold's strategy reaches the ground state in a few steps.

\section{Conclusions}
The present paper introduced optimal flipping policies for kinetic Ising models, using the stochastic Bellman equation as a prototypical Maxwell demon eager to extract the energy from a thermal system. As we have shown in this paper, these demons have less power than a typical Maxwell demon, as they can only operate locally on a single spin, and have to satisfy the detailed balance. They thus parallel the typical strategy of a Monte Carlo algorithm, with the different that instead than thermalizing the model, their intent is lowering the energy given the Markov chain transition probabilities. As we have shown, their strategy and approach strongly depends on how fast they want to extract energy from the system, changing from a greedy to a wise approach when discounting is high or low respectively. Unfortunately, the Bellman equation still is plagued by the curse of dimensionality, and only small systems could be analyzed. Nonetheless, we have shown that these present interesting strategy changes also for small systems.

This approach can also be interpreted as the optimal sequential strategy to optimize Ising models in an uncertain but time-invariant environment. As a way to overcome the curse of dimensionality, we note that this is exactly the same problematic that reinforcement learning aims to tackle, which is the natural extension of this work \cite{rl}. In fact, in reinforcement learning agents have ``imperfect information", and do not necessarily know fully the state space, and in our case the energy landscape. In this sense, a natural extension of our work is the one in which a kobold is ``trained" via a reinforcement learning algorithm, learning the strategy by testing the results of actions on the phase space, and updating iteratively the strategy. This will be the focus of future works.



\acknowledgments
 This work was carried out under the auspices of the NNSA of the U.S. DoE at LANL under Contract No. DE-AC52-06NA25396, and in particular grant PRD20190195 from the LDRD. This paper was inspired by ``Meister Eder und sein Pumuckl".

\end{document}